\documentclass[11pt]{article}
\usepackage{moriond}
\usepackage{amsmath,amssymb}

\def\be{\begin{equation}}
\def\ee{\end{equation}}
\def\bea{\begin{eqnarray}}
\def\eea{\end{eqnarray}}

\def\SU{{\rm SU}}

\newcommand{\Photo}{\bigskip\includegraphics[height=35mm]{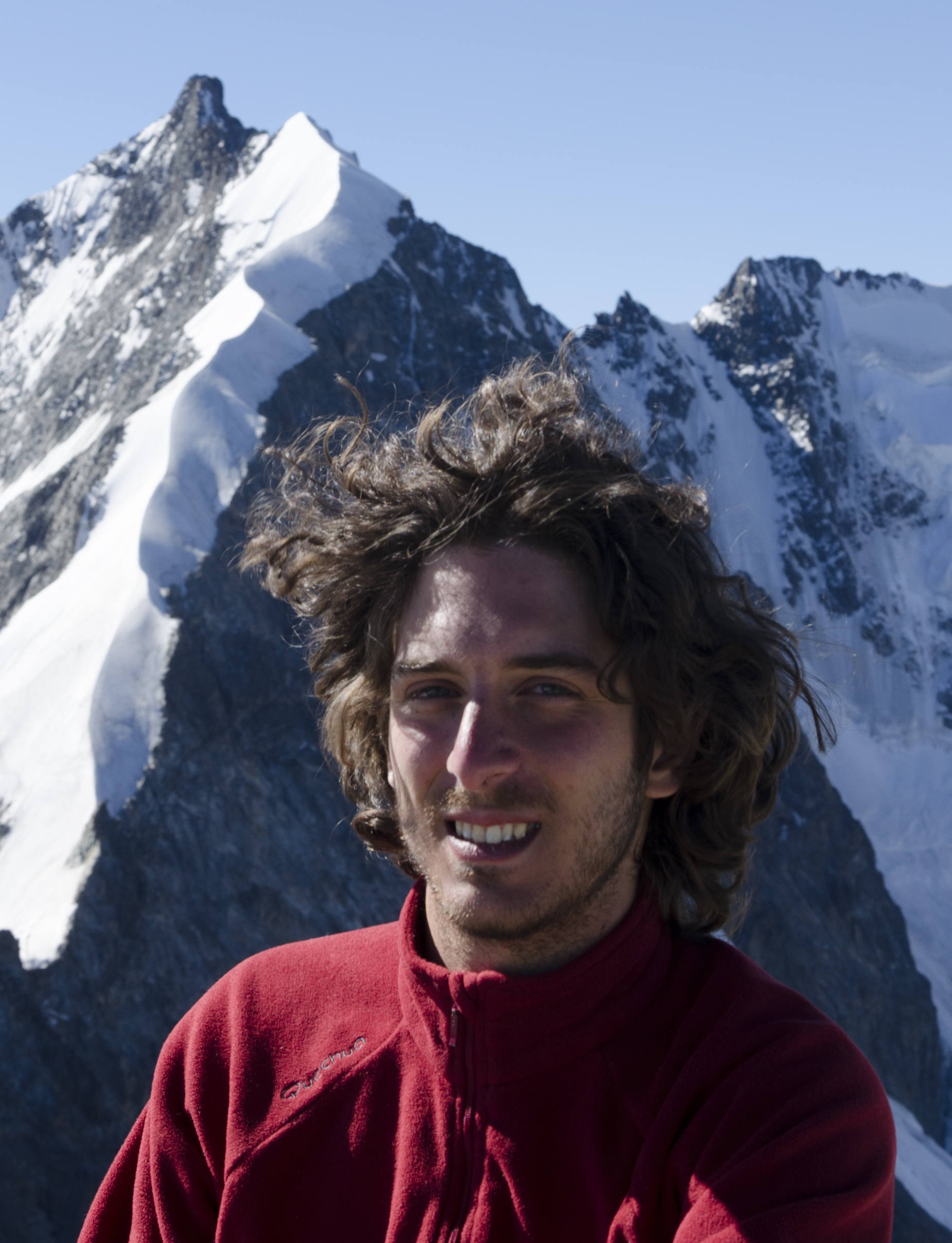}}

\begin{document}
FLAVOUR(267104)-ERC-73
\vspace*{4cm}
\title{NATURAL SCALARS IN THE NMSSM}

\author{ DARIO BUTTAZZO }

\address{TUM Institute for Advanced Study, Lichtenbergstr. 2a, 85747 Garching, Germany}

\maketitle\abstracts{
In the motivated hypothesis that the scalar bosons of the Next-to-Minimal Supersymmetric Standard Model (NMSSM) be the lightest new particles around, a possible strategy to search for signs of the extra CP-even states is outlined.
It is shown how the measurements of the couplings of the 126 GeV Higgs boson constrain the region of the physical parameters in a generic NMSSM which minimises the fine-tuning of the electroweak scale. We also determine the cross section for the production of a heavier CP-even scalar, together with its most relevant branching ratios.}

\section{Introduction}

Is the Higgs boson recently discovered at the LHC alone, or is it a member of an extended family of scalar particles? While this is a relevant question on its own, it has a fundamental importance in the context of supersymmetry, where at least a second doublet is required.

Also in view of the negative results in the searches for many supersymmetric partners up to TeV-range masses, the extra Higgs bosons may well be the lightest particles in the spectrum -- perhaps with the exception of the LSP -- making the search for these scalar states an important task for present and future experiments.

The Higgs system of the NMSSM contains two scalar $\SU(2)_L$ doublets $H_u$ and $H_d$, and a complex singlet $S$, all parts of the corresponding chiral supermultiplets, coupled through a cubic term $\lambda H_u H_d S$ in the superpotential.~\cite{Fayet}
Taking naturalness as a guideline, there are two main reasons for considering the NMSSM:~\cite{Barbieri} 
\begin{enumerate}
\item it adds a new tree-level contribution to the Higgs mass, which makes a 126 GeV Higgs boson compatible with lighter stops with respect to the MSSM;
\item it reduces the fine-tuning of the electroweak scale $v$ (fixed by the weak gauge couplings in the MSSM) for moderate $\tan\beta$ and $\lambda\approx 1$.
\end{enumerate}

The aim here is to present an analytical study of the Higgs system of the general NMSSM in a most natural scenario, without specifying any particular form of the scalar potential and avoiding the use of benchmark points.~\footnote{See Barbieri {\it et al.} \cite{Barbieri1,Barbieri2} for more details.}
In order to simplify the analysis, however, we shall make the following motivated assumptions:
\begin{enumerate}
\item we assume a negligibly small CP violation in the Higgs system, and therefore we ignore the two CP-odd states, which do not influence the physics of the CP-even states;
\item we neglect any effect from loops of supersymmetric particles other than the correction $\Delta_t$ to the quartic coupling of $H_u$ due to the top-stop loop; this is motivated by the choice of a spectrum with all the s-particles as close as possible to their ``naturalness limit'';
\item 
we assume $\mu A_t \lesssim m_{\tilde t}^2$, due to naturalness arguments, where $m_{\tilde t}$ is the average stop mass, $A_t$ is its trilinear coupling, and $\mu$ is the quadratic term in the superpotential;
\item we do not include any invisible decay of the lightest Higgs boson -- e.g. into a pair of neutralinos; this can easily be corrected for rescaling all the branching ratios and signal strengths by a common factor $\Gamma/(\Gamma + \Gamma_{\rm inv})$.
\end{enumerate}

\section{Parameter space of a generic NMSSM}

Assuming a negligibly small violation of CP in the Higgs sector, the three neutral CP-even fields $\mathcal{H} = (H_u^0,H_d^0,S)^\mathsf{T}$ are related to the physical mass eigenstates $\mathcal{H}_{\rm ph} = (h_1,h_2,h_3)^\mathsf{T}$ by
\begin{equation}\label{rotation}
\mathcal{H} = R_\alpha^{12}R_\gamma^{23}R_\sigma^{13}\mathcal{H}_{\rm ph} \equiv R\,\mathcal{H}_{\rm ph},
\end{equation}
where $R_x^{ij}$ are rotations by an angle $x$ in the $(i,j)$ sector. Their squared mass matrix reads, in the $\mathcal{H}$ basis,
\begin{equation}\label{scalar_mass_matrix}
{\cal M}^2=\left(
\begin{array}{ccc}
m_A^2 c^2_\beta+m_Z^2 s^2_\beta +\Delta_t^2/s_\beta^2 & \left(2 v^2 \lambda ^2-m_A^2-m_Z^2\right) c_\beta s_\beta &  v M_1  \\
 \left(2 v^2 \lambda ^2-m_A^2-m_Z^2\right) c_\beta s_\beta & m_Z^2 c^2_\beta + m_A^2 s^2_\beta &  v M_2  \\
  v  M_1 &  v M_2 & M_3^2
\end{array}
\right),
\end{equation}
where
\begin{equation}
\label{mHcharged}
m_A^2 = m_{H^{\pm}}^2 - m_W^2 +\lambda^2 v^2,
\end{equation}
$m_{H^{\pm}}$ is the physical mass of the single charged Higgs boson, $\Delta_t^2$ is the well-known effect of the top-stop loop corrections to the quartic coupling of $H_u$, and $v \simeq 174$ GeV. Here and in the following we write $s_x = \sin x, c_x = \cos x$.
We neglect the corrections to $\mathcal{M}_{11}$ and $\mathcal{M}_{12}$, which are suppressed as the second and first power of $\mu A_t/m_{\tilde t}^2$, respectively, and the analogous correction to \eqref{mHcharged}.
We leave unspecified the other parameters $M_1, M_2, M_3$ in \eqref{scalar_mass_matrix}, which are not directly related to physical masses and  depend on the particular NMSSM under consideration -- i.e. the form of the singlet potential.
The matrix $\mathcal{M}$ is related to the physical scalar masses by
\begin{equation}
R^\mathsf{T} {\cal M}^2 R = \mathrm{diag}(m_{h_1}^2, m_{h_2}^2, m_{h_3}^2).
\label{diag_matrix}
\end{equation}

In the following, we identify $h_1$ with the state found at the LHC, so that $m_{h_1} = 125.7$ GeV. For simplicity we shall always consider 
$h_1$ as the lightest CP-even state, although other cases with a lighter scalar are still compatible with current data.~\cite{Barbieri2}

Although the full matrix $\mathcal{M}^2$ depends on the specific model in consideration, its $2\times 2$ submatrix in the (1,2) sector provides, by the use of \eqref{diag_matrix}, three relations between the mixing angles and the physical masses \cite{Barbieri1,Barbieri2} which do not depend on the unknown quantities $M_1$, $M_2$, $M_3$. The Higgs system of the NMSSM is thus completely determined by the parameters $m_{h_{1,2,3}}$, $m_{H^\pm}$, $\lambda$, $t_\beta$, $\Delta_t$.
Due to the large number of free parameters, in order to simplify the analysis we will consider the limiting cases where only two states out of three are light:
\begin{itemize}
\item Singlet decoupled: $m_{h_3} \gg m_{h_1}, m_{h_2}$ or $M_3^2 \gg v M_1, v M_2$, and $\sigma, \gamma \rightarrow 0$;
\item Doublet decoupled: $m_{h_2} \gg m_{h_1}, m_{h_3}$ or $m_A^2 \gg v M_1, v M_2$, and $\sigma, \delta = \alpha - \beta + \pi/2 \rightarrow 0$.
\end{itemize}

\begin{figure}
\begin{center}
\includegraphics[width=.49\textwidth]{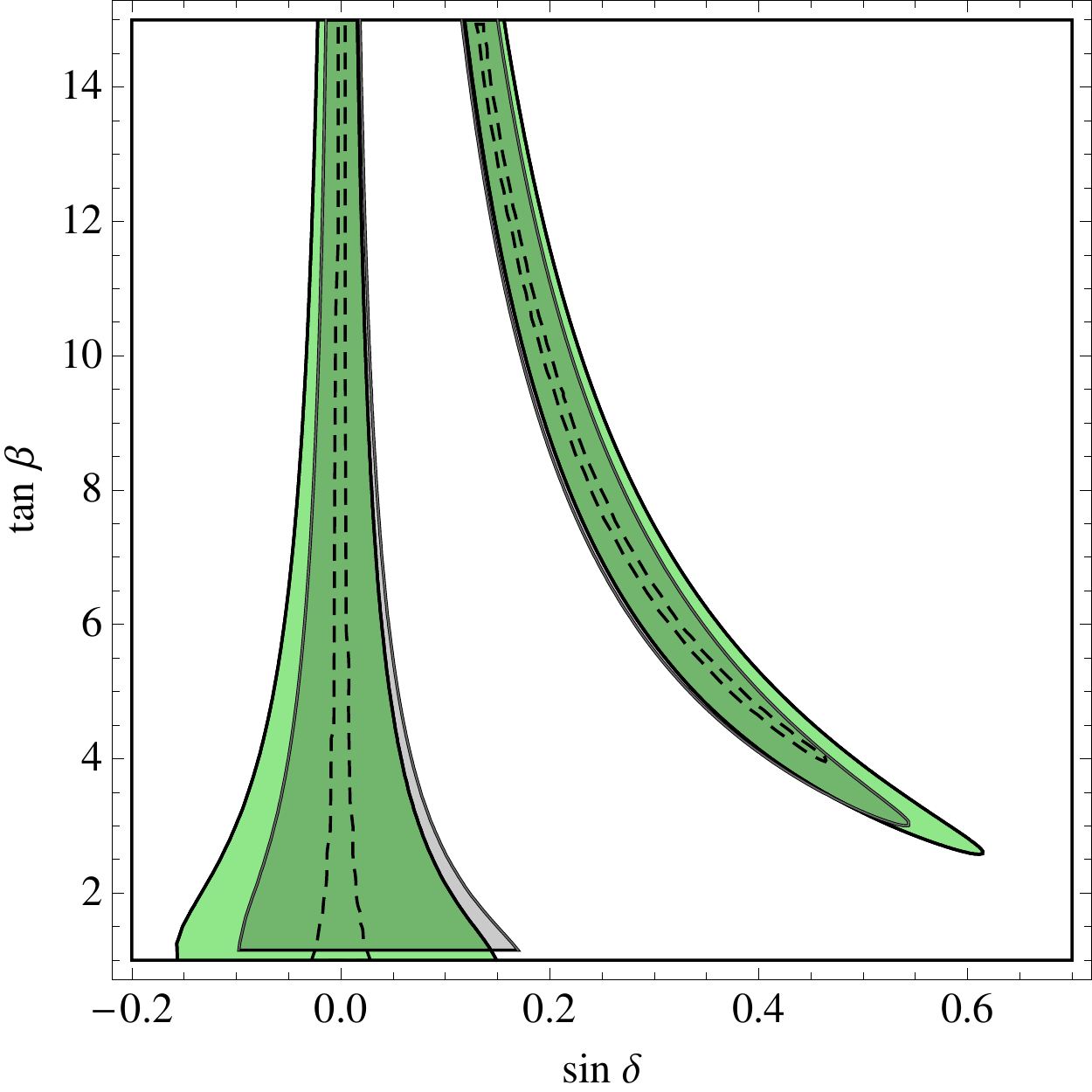}\hfill
\includegraphics[width=.49\textwidth]{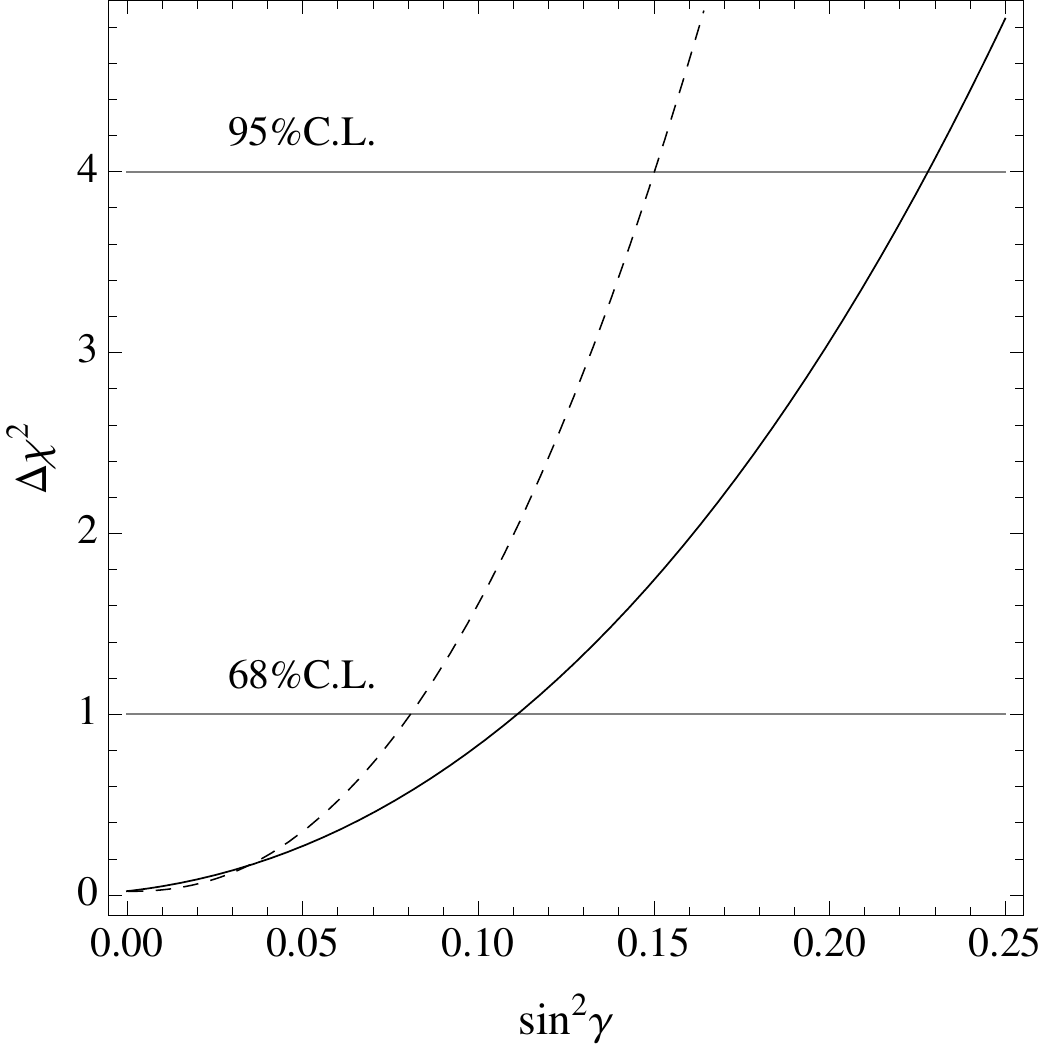}
\end{center}
\caption{\label{fit} Fit of the measured signal strengths of $h_1 = h_{\text{LHC}}$. Left: 3-parameter fit of $t_\beta$, $s_{\delta}$ and $s_{\gamma}^{2}$. The allowed regions at 95\% C.L. are given for $s_{\gamma}^{2} = 0$ (green) and $s_\gamma^2 = 0.15$ (grey). Note that the two regions overlap in part. The dashed line shows the fit reach expected at LHC14 for $s_\gamma^2 = 0$. Right: fit of $s_{\gamma}^{2}$ in the case of $\delta = 0$ (solid) and its projection at LHC14 (dashed).}
\end{figure}

\section{Higgs couplings}

\begin{table}[b]
\caption{Projected uncertainties of the measurements of the signal strengths of $h_{\rm LHC}$, normalized to the SM, at the 14 TeV LHC with 300 fb$^{-1}$, both for ATLAS and CMS.}
\label{table}
\vspace{0.4cm}
\begin{center}
\begin{tabular}{|l|cc|}
\hline
& ATLAS & CMS \\
\hline
$h\to \gamma\gamma$ & 0.16 & 0.15 \\
$h\to ZZ$ & 0.15 & 0.11 \\
$h\to WW$ & 0.30 & 0.14 \\
$Vh\to V b\bar b$ & -- & 0.17 \\
$h\to \tau\tau$ & 0.24 & 0.11 \\
$h\to \mu\mu$ & 0.52 & -- \\
\hline
\end{tabular}
\end{center}
\end{table}

From \eqref{rotation}, $h_1 = h_{\text{LHC}}$ is related to the gauge eigenstates by
\begin{equation}\label{rotation_h1}
h_1 = c_{\gamma} (-s_{\alpha} H_d^0 + c_\alpha H_u^0) + s_{\gamma} S,
\end{equation}
and similar relations, also involving the angle $\sigma$, hold for $h_2$ and $h_3$.
The angles $\delta = \alpha - \beta + \pi/2$ and $\gamma$ alone thus determine the couplings of $h_1$ to the fermions and to vector boson pairs, normalised to the corresponding couplings of the SM Higgs boson,
\begin{align}
\frac{g_{h_1tt}}{g^{\text{SM}}_{htt}} &= c_\gamma\Big(c_\delta +\frac{s_\delta}{t_\beta}\Big), & \frac{g_{h_1bb}}{g^{\text{SM}}_{hbb}} &= c_\gamma(c_\delta -s_\delta t_\beta ), & \frac{g_{h_1VV}}{g^{\text{SM}}_{hVV}} &=  c_\gamma c_\delta.
\label{h1couplings}
\end{align}

A fit of all ATLAS, CMS and TeVatron data collected so far~\cite{rates}\textsuperscript{,}~\footnote{See Barbieri {\it et al.} \cite{Barbieri1,Barbieri2} for a detailed list of references.} on the various signal strengths of $h_{\text{LHC}}$ can then be used to put bounds on $\delta$ and $\gamma$ (as a function of $t_\beta$). We perform this fit adapting the code of Giardino {\it et al.}~\cite{Giardino} As stated above, we do not include in \eqref{h1couplings} and in the fit any loop effect from supersymmetric particles. The 95\% C.L. allowed regions for $\delta$, at different fixed values of $\gamma$, and for $\gamma$ at $\delta = 0$, are shown in Figure~\ref{fit}.

To quantify the impact of the future experimental improvement in the measurements of the various signal strengths of $h_{\rm LHC}$, as foreseen in the next run of the LHC, we have repeated the previous fit assuming the expected errors on the signal strengths~\cite{projection} with a luminosity of 300 fb$^{-1}$ at $\sqrt{s} = 14$ TeV as in Table~\ref{table}, and central values as in the Standard Model. The results of this projection are also shown in Figure~\ref{fit} as a comparison. Notice that, while a big improvement is expected in the -- already quite precise -- determination of $\delta$, the fit of $\gamma$ will only marginally improve.

\section{Singlet decoupled}
If the singlet is decoupled, the only nonzero mixing angle is
\be
s_\delta^2 = \frac{m_{hh}^2 - m_{h_1}^2}{m_{h_2}^2 - m_{h_1}^2},
\ee
where
\be
\label{mhh}
m_{hh}^2 = m_Z^2 c_{2\beta}^2 + v^2\lambda^2s_{2\beta}^2 + \Delta_t^2,
\ee
and both $m_{h_2}$ and $m_{H^\pm}$ can be expressed in terms of the three parameters $\lambda, t_\beta, \Delta_t$. The $\lambda\to 0$ limit corresponds to the MSSM case. The dependence on $\Delta_t$ is very mild and can be neglected if $\Delta_t$ itself is not too large, corresponding to a moderate level of fine-tuning. We therefore choose to fix $\Delta_t = 75$ GeV, which is compatible with an average stop mass of about 700~GeV.

\begin{figure}
\centering%
\includegraphics[width=0.49\textwidth]{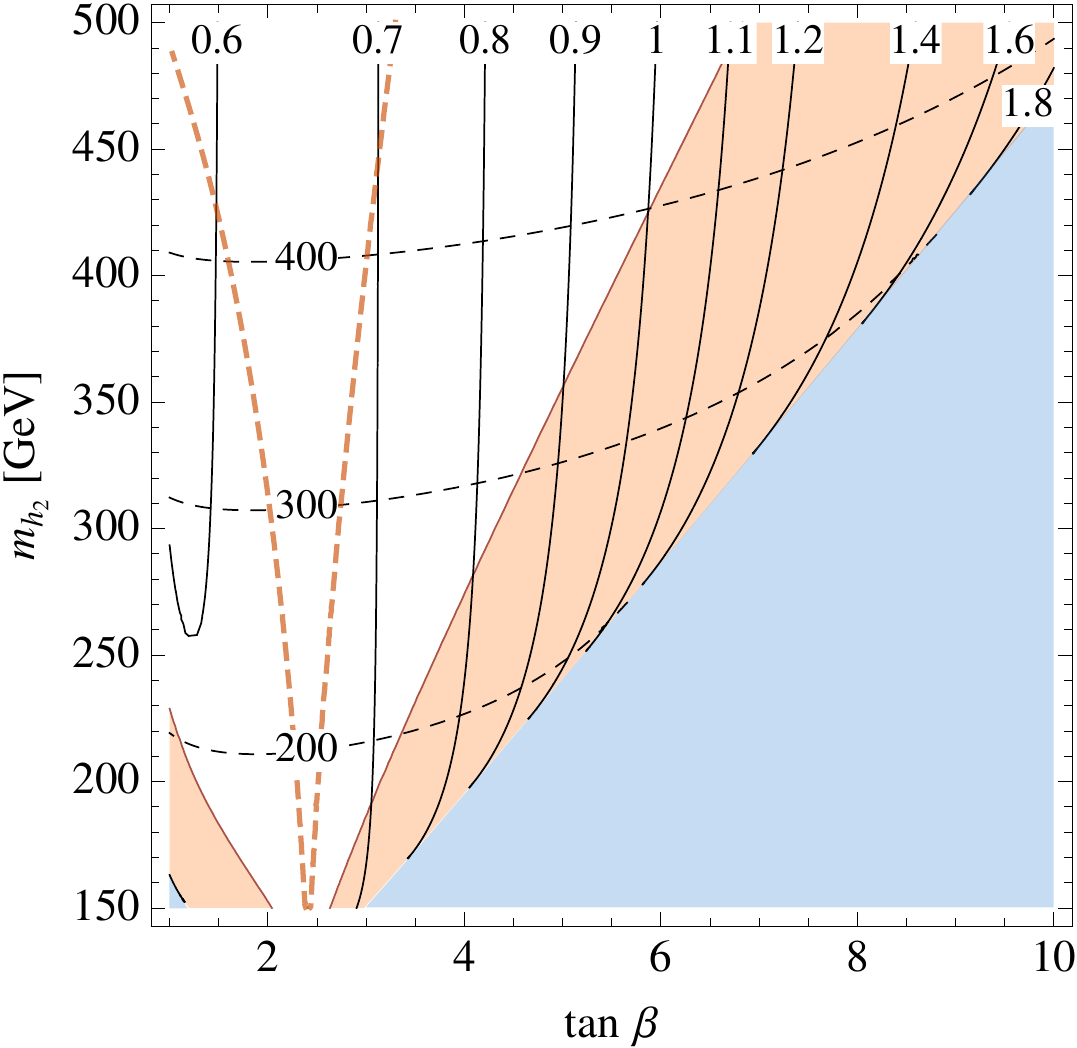}\hfill
\includegraphics[width=0.49\textwidth]{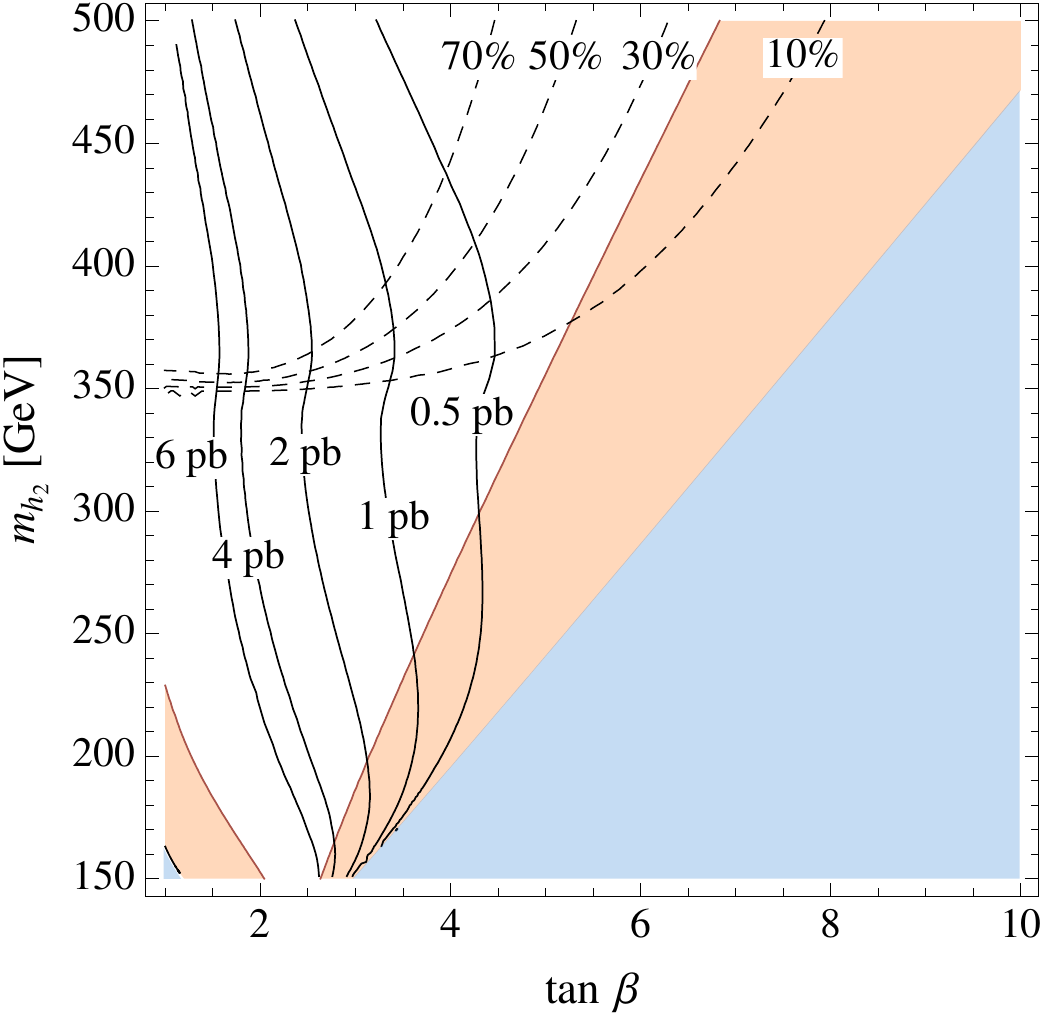}
\caption{\label{Sdec}Singlet decoupled in the $(t_\beta, m_{h_2})$ plane. The orange region is excluded at 95\% C.L. by the Higgs fit, the blue region is unphysical. Left: the colored dashed lines show the fit reach expected at LHC14, black isolines of $\lambda$ (solid) and $m_{H^\pm}$ (dashed). Right: isolines of the gluon fusion production cross-section $\sigma(gg\to h_2)$ at 14~TeV (solid) and of the decay branching ratio into top quark pairs ${\rm BR}(h_2\to t\bar t)$ (dashed).}
\end{figure}

The left panel of Figure~\ref{Sdec} shows the regions excluded by the fit, present and forseen, in the $(t_\beta,m_{h_2})$ plane, together with the isolines of $\lambda$ and $m_{H^\pm}$. A second light state is allowed for small $t_\beta$ and moderate values of $\lambda\lesssim 1$, unlike in the MSSM,~\cite{Barbieri1} although masses below 300~GeV are disfavoured by the presence of a too light charged Higgs. Notice that with the improved measurements of the signal rates of $h_{\rm LHC}$ in the next stage of the LHC it will be possible to probe a large fraction of the parameter space solely by the Higgs fit.

The couplings of $h_2$, on the other hand, are given by
\begin{align}
\frac{g_{h_2 tt}}{g_{htt}^{\rm SM}} &= s_\delta - \frac{c_\delta}{t_\beta}, & \frac{g_{h_2 bb}}{g_{hbb}^{\rm SM}} &= s_\delta + c_\delta t_\beta, & \frac{g_{h_2 VV}}{g_{hVV}^{\rm SM}} &= s_\delta,
\end{align}
and allow to calculate its production cross-sections and most relevant branching ratios. The small values of $\lambda$ in the region allowed by the fit make the phenomenology of $h_2$ quite similar to the one of the heavier Higgs state of the MSSM. Its dominant decay mode is into fermions, either top or bottom quarks, depending on the mass $m_{h_2}$ and on $t_\beta$. The right panel of Figure~\ref{Sdec} shows the predictions for the NNLL gluon fusion production cross-section of $h_2$ at $\sqrt{s} = 14$ TeV, and for its branching ratio into top quark pairs ${\rm BR}(h_2\to t\bar t)$, in the same $(t_\beta, m_{h_2})$ plane.

\section{Doublet decoupled}

\begin{figure}
\centering%
\includegraphics[width=0.49\textwidth]{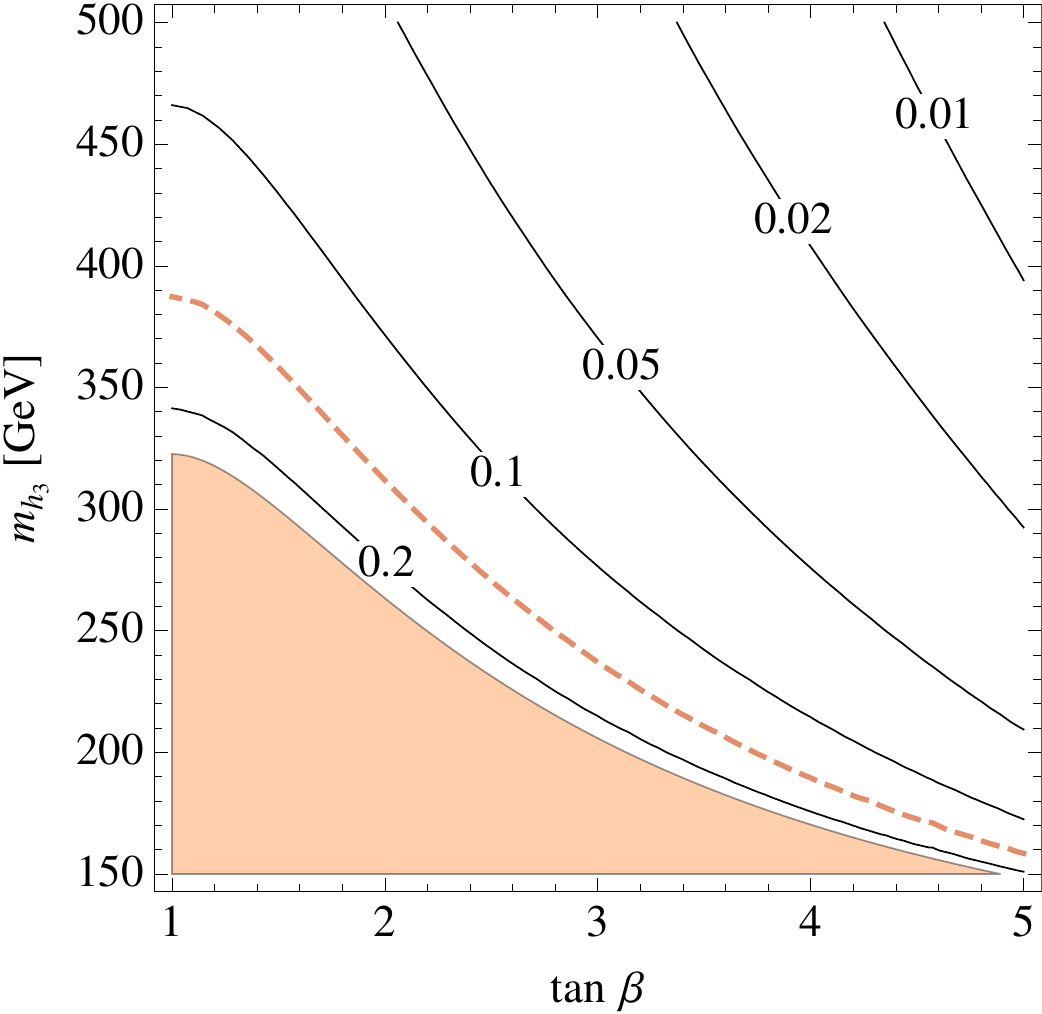}\hfill
\includegraphics[width=0.49\textwidth]{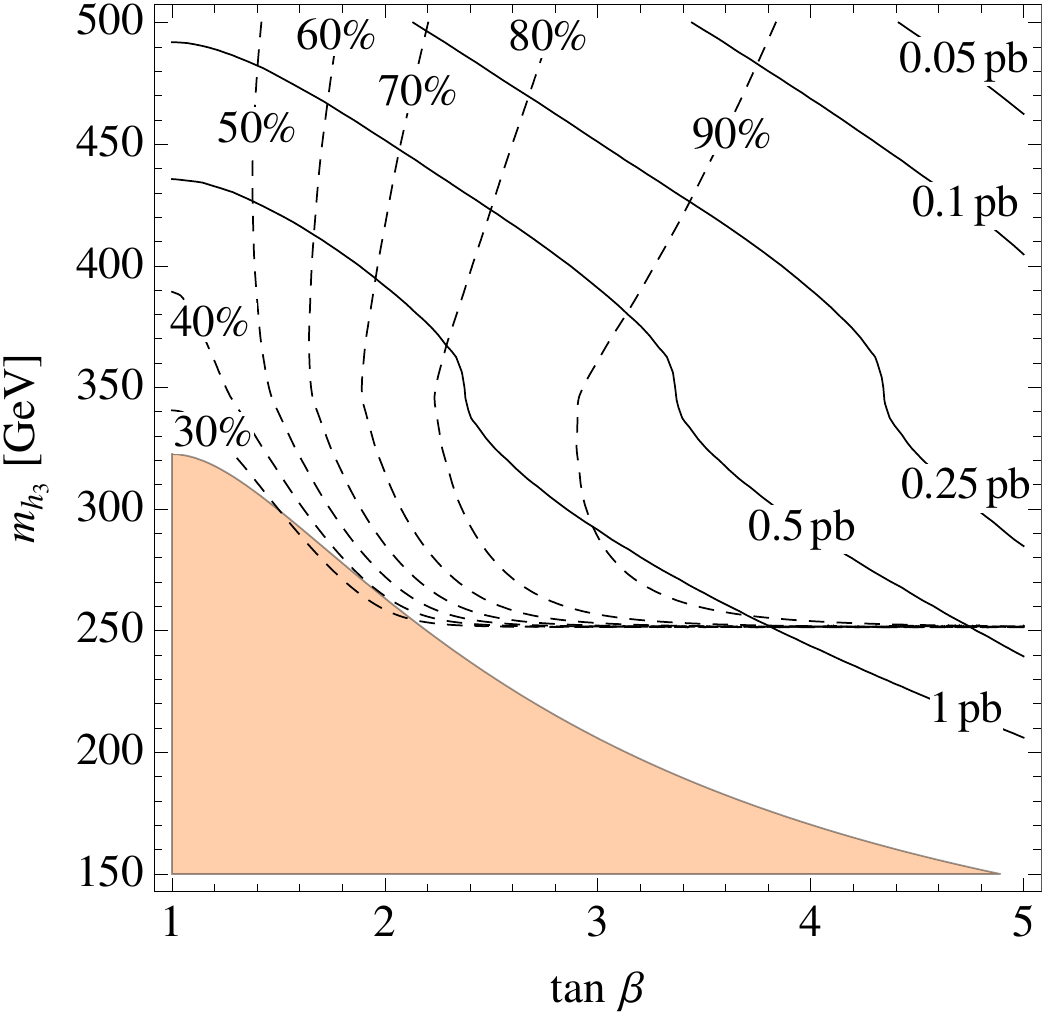}
\caption{\label{Hdec}Doublet decoupled in the $(t_\beta, m_{h_3})$ plane, for fixed $\lambda = 1$. The orange region is excluded at 95\% C.L. by the Higgs fit. Left: the coloured dashed line shows the fit reach expected at LHC14, black isolines of $s_\gamma^2$. Right: isolines of the gluon fusion production cross-section $\sigma(gg\to h_3)$ at 14 TeV (solid) and of the decay branching ratio of $h_3$ into two Higgs bosons ${\rm BR}(h_3\to h_1 h_1)$, with $v_S = 2v$ (dashed).}
\end{figure}

If the standard-like Higgs mixes with a singlet, and the second doublet is decoupled, the only nonzero mixing angle is
\be
s_\gamma^2 = \frac{m_{hh}^2 - m_{h_1}^2}{m_{h_3}^2 - m_{h_1}^2}.
\ee
Since the off-diagonal entries $M_1$ and $M_2$ in \eqref{scalar_mass_matrix} are unknown, there is one free parameter more respect to the previous case. We therefore fix $\lambda = 1$, along with choosing $\Delta_t = 75$ GeV as before, in order to produce the plots of Figure~\ref{Hdec}. The left panel shows, as before, the 95\% C.L. excluded region from the Higgs fit, together with its projection at 14 TeV, and the isolines of the mixing angle $s_\gamma^2$. Recall from \eqref{mhh} that $s_\gamma^2$ increases with $\lambda$, thus enlarging the region allowed by the fit for smaller values of $\lambda$.~\cite{Barbieri1} The mild improvement in the fit foreseen for the next run of the LHC in this particular case makes the direct searches for the heavy state $h_3$ crucial.

The couplings of $h_3$, due to its singlet-like nature, are proportional to the ones of a standard Higgs boson with mass $m_{h_3}$,
\be
\frac{g_{h_3 tt}}{g_{htt}^{\rm SM}} = \frac{g_{h_3 bb}}{g_{hbb}^{\rm SM}} = \frac{g_{h_3 VV}}{g_{hVV}^{\rm SM}} = -s_\gamma.
\ee
The branching ratios of $h_3$ are therefore the same as the standard ones, while the production cross-sections and decay widths are simply rescaled by $s_\gamma^2$ (still neglecting radiative corrections from supersymmetric particles). The dominant decay channel, if kinematically allowed, is the one of $h_3$ into two $h_1$ bosons, followed in importance by the one into two vector bosons. Unfortunately, all the triple scalar couplings depend on the particular form of the singlet potential, and are therefore model dependent. In the limit of large $\lambda$, however, this dependence can be parametrized simply by the vacuum expectation value of the singlet $v_S$.~\cite{Barbieri1}

In the right panel of Figure~\ref{Hdec} we show the branching ratio ${\rm BR}(h_3\to h_1 h_1)$ for $v_S = 2 v$, together with the gluon fusion production cross-section at $\sqrt{s} = 14$ TeV. Although not easily, searches in the $b\bar b\gamma\gamma$ and $b\bar b b\bar b$ channels could reach comparable sensitivities in the near future, probing regions of the parameter space difficultly accessible by other means.

Finally, it is worth to mention that very large deviations of the triple $h_1$ coupling from the standard value can arise in some part of the parameter space,~\cite{Barbieri1} perhaps making this measure also accessible at the LHC in the future.

\section{Conclusions}
We have analysed the Higgs system of a most natural NMSSM, focussing on relations between physical parameters. The modified couplings of $h_1$, which influence the signal strengths measured at the LHC, provide a powerful tool to exclude regions of the parameter space.

We have considered two limiting cases in which one of the states is much heavier than the others. In both cases a second light neutral CP-even scalar is consistent with all the constraints.

If the singlet is decoupled, it will be possible to thoroughly explore the parameter space combining the refined measurements of the signal strengths of $h_{\rm LHC}$ with searches for the second Higgs decaying into fermions in the remaining allowed regions.

The case where the doublet is decoupled is more difficult to test. The $h_1$ signal strengths are not very sensitive to the mixing with the second state, due to its singlet-like nature. Furthermore, the small production cross-section of $h_3$, together with its large branching ratio into a pair of $h_1$, whenever kinematically allowed, makes the search at the LHC challenging, although not impossible. 

It will in any case be interesting to follow the progression of the experimental searches for additional Higgs-like states, either direct or indirect, which are an independent way to probe weak-scale supersymmetry, complementary to the search for superpartners.

\section*{Acknowledgments}

I would like to thank Riccardo Barbieri, Kristjan Kannike, Filippo Sala and Andrea Tesi for the fruitful collaborations, and the organizers of the Rencontres de Moriond for the pleasant and stimulating environment. Thanks also to Maxime Gouzevitch and Caterina Vernieri for interesting discussions. This work was supported by the ERC Advanced Grant project {\sc Flavour} (agreement n. 267104).


\section*{References}

\begin{thebibliography}{99}
\bibitem{Fayet}P. Fayet, {\it Nucl.~Phys.~B} {\bf 90} (1975) 104-124.

\bibitem{Barbieri}R. Barbieri, L. J. Hall, Y. Nomura and V. S. Rychkov, {\it Phys.~Rev.~D} {\bf 75} (2007) 035007, \href{http://arxiv.org/abs/hep-ph/0607332}{arXiv:hep-ph/0607332};
L. J. Hall, D. Pinner and J. T. Rudermann, {\it JHEP} {\bf 1204} (2012) 131, \href{http://arxiv.org/abs/1112.2703}{arXiv:1112.2703 [hep-ph]}.

\bibitem{Barbieri1}R. Barbieri, D. Buttazzo, K. Kannike, F. Sala and A. Tesi, {\it Phys.~Rev.~D} {\bf 87} (2013) 115018, \href{http://arxiv.org/abs/1304.3670}{arXiv:1304.3670 [hep-ph]}.

\bibitem{Barbieri2}R. Barbieri, D. Buttazzo, K. Kannike, F. Sala and A. Tesi, {\it Phys.~Rev.~D} {\bf 88} (2013) 055011, \href{http://arxiv.org/abs/1307.4937}{arXiv:1307.4937 [hep-ph]}.

\bibitem{Giardino}P. P. Giardino, K. Kannike, I. Masina, M. Raidal and A. Strumia, \href{http://arxiv.org/abs/1303.3570}{arXiv:1303.3570[hep-ph]}.

\bibitem{rates}ATLAS Collaboration, \href{https://twiki.cern.ch/twiki/bin/view/AtlasPublic/HiggsPublicResults}{twiki.cern.ch/twiki/bin/view/AtlasPublic/HiggsPublicResults};\\ CMS Collaboration, \href{https://twiki.cern.ch/twiki/bin/view/CMSPublic/PhysicsResultsHIG}{twiki.cern.ch/twiki/bin/view/CMSPublic/PhysicsResultsHIG}.

\bibitem{projection}ATLAS Collaboration, \href{http://cds.cern.ch/record/1484890}{ATL-PHYS-PUB-2012-004}; CMS Collaboration, \href{http://cds.cern.ch/record/1494600}{CMS-NOTE-2012-006}.

\end{thebibliography}
\end{document}